# A cluster partitioning method: determination of density matrices of solids and comparison with X-ray experiments

Sébastien RAGOT[#], Jean-Michel GILLET[#], Pierre J. BECKER[#@]

[#] Laboratoire Structure, Propriété et Modélisation des Solides (CNRS, Unité Mixte de Recherche 85-80). École Centrale Paris, Grande Voie des Vignes, 92295 CHATENAY-MALABRY, FRANCE

[@] Université Catholique de l'ouest. 1, place André-Leroy. BP808, 49008 ANGERS Cedex 01, FRANCE

## Abstract

In this paper we show that 1-electron properties such as Compton profiles and structure factors of crystals can be asymptotically retrieved through cluster-based calculations, followed by an appropriate partition of the 1-electron reduced density matrix (1RDM). This approach, conceptually simple, is checked with respects to both position and momentum spaces simultaneously for insulators and a covalent crystal. Restricting the calculations to small clusters further enables a fair description of local correlation effects in ionic compounds, which improves both Compton profiles and structure factors *vs.* their experimentally determined counterparts.

# I. Introduction

Electron correlation, and its effect on cohesive properties, has been a subject of constant research during the last decades[1]. The development of density functional theory (DFT) constitutes a major progress in this field[2]. However, it can sometimes be misleading since based on the only charge density $\rho(r)$: momentum space properties are for instance driven by the non-diagonal part of the exactly $N$-representable 1RDM[2]. Given the arbitrariness of Kohn-Sham orbitals for considerations in momentum space, it is more convenient to deal with an $N$-representable 1RDM.

The question is then: « how can we simulate in a simple manner an ion or a molecule in a solid medium at correlated level, with a model valid in both position and momentum spaces (*i.e.*, $R$ and $P$-spaces)? ». A previously elaborated model, based on repeated clusters[3], turned out to be efficient for simple ionic compounds[3,31]. However, such a model is rather difficult to extend to complex systems. We therefore propose another approach in which we make use of $N$-representable 1RDMs and standard *ab initio* codes. One very common possibility relies on performing cluster-based calculations, though the correlation problem for insulators and semi-conductors can be handled in many different ways[4,5,6]. A cluster approach might indeed lead to a charge density at the center of the cluster that is close to the crystal one but it cannot permit, however, to isolate the momentum density characteristic of that of the solid, due to surface effects. Therefore, we make use of a simple partition scheme in order to ovoid this shortcoming (section II). We show in the third section that, when restricting at Hartree-Fock (HF) level, the results converge towards crystal ones for some insulators and for silicon. Using small clusters further allows for a fair estimation of local correlation effects (section IV) in both LiH and MgO crystals. The correlation corrections are consistent with experimental deviations to HF Compton profiles and structure factors.

## II. A cluster partitioning method (CPM)

In the case of an isolated molecule, treated within the Born-Oppenheimer approximation, the 1RDM writes as $\rho(r,r') = \sum_{A,B} \sum_{i \in A, j \in B} c_{ij}^{AB} \varphi_{iA}(r - R_A) \varphi_{jB}^*(r' - R_B)$. Atomic orbitals $\varphi_{iA}$ are centered on $R_A$ (pointing at the center of an atom) and are assumed to be real, for simplicity. Note that the 1RDM can formally be rewritten as

$$\rho(r,r') = \sum_{A,B} \rho_{A,B}(r - R_A, r' - R_B) \tag{1}$$

Separating 1 and 2 center terms in (1) and symmetrizing them afterwards gives the following formulation of the 1RDM

$$\rho = \sum_A \left\{ \rho_{A,A} + \tfrac{1}{2} \sum_{B(\neq A)} [\rho_{A,B} + \rho_{B,A}] \right\} \tag{2}$$

This partition scheme is, so far, nothing else than a Mulliken-like partition scheme. Dropping the explicit dependence on $R_B$ atomic positions, we can rewrite the 1RDM as $\rho(r,r') = \sum_A \rho_A(r - R_A, r' - R_A)$. Conversely, we might write $\rho(r,r')$ in its intracular-extracular representation[7]:

$$\rho(r,r') \equiv \tilde{\rho}(R,s) = \sum_A \tilde{\rho}_A(R - R_A, s) \tag{3}$$

where $R$ stands for $(r + r')/2$ and $s$ is the difference vector $r - r'$. Other partition schemes could obviously result in a 1-center decomposition of the 1RDM similar to (3).

When extending the molecule to a crystal with a group of $N$ atoms as a unit basis, $\tilde{\rho}(R,s)$ becomes invariant by a translation $L$ (a lattice vector) of the $R$ coordinate:

$$\tilde{\rho}(R,s) = \sum_L \sum_{A=1}^N \tilde{\rho}_A(R - L - R_A, s) = \sum_L \tilde{\rho}_L(R - L, s) \tag{4}$$

The momentum density is defined as

$$n(p) = \frac{1}{(2\pi)^3} \int \rho(r,r') e^{ip \cdot (r-r')} dr dr' = \frac{1}{(2\pi)^3} \int \tilde{\rho}(R,s) e^{ip \cdot s} dR ds \tag{5}$$

Thus, $n(\boldsymbol{p})$ turns out to be the Fourier transform of the so-called auto correlation function[8], which is obtained by integrating the 1RDM over the $\boldsymbol{R}$ coordinate: $B(\boldsymbol{s}) = \int \tilde{\rho}(\boldsymbol{R},\boldsymbol{s})d\boldsymbol{R}$. Finally, each basis unit term $\tilde{\rho}_L(\boldsymbol{R}-\boldsymbol{L},\boldsymbol{s})$ from (4) gives an identical contribution to $n(\boldsymbol{p})$ and $B(\boldsymbol{s})$, so that the arbitrariness of the 1RDM partition disappears and only one term is to be computed. The impulse Compton profile[9] $J(p_z)$ is then obtained from

$$J(p_z) = \int n(\boldsymbol{p})\delta(p_z - q)d\boldsymbol{p} = \frac{1}{2\pi}\int B(0,0,s_z)e^{ip_z s_z}dp_z \tag{6}$$

Practically, two-center contributions in (2) decrease as overlaps between orbitals become negligible. For an insulator $AB$ with NaCl-like structure, two cluster calculations can be performed for both $A$ and $B$ ions, each located at the center of its corresponding cluster, in order to preserve the local symmetry of the environment. Electronic densities can then be recovered by summing the contributions of each central ion:

$$\tilde{\rho}_A = \tilde{\rho}_{A,A} + \tfrac{1}{2}\sum_{C(\neq A)}\left[\tilde{\rho}_{A,C} + \tilde{\rho}_{C,A}\right]$$

$$\tilde{\rho}_B = \tilde{\rho}_{B,B} + \tfrac{1}{2}\sum_{C(\neq B)}\left[\tilde{\rho}_{B,C} + \tilde{\rho}_{C,B}\right] \tag{7}$$

$$\tilde{\rho}_L(\boldsymbol{R}-\boldsymbol{L},\boldsymbol{s}) = \tilde{\rho}_A(\boldsymbol{R}-\boldsymbol{L}-\boldsymbol{R}_A,\boldsymbol{s}) + \tilde{\rho}_B(\boldsymbol{R}-\boldsymbol{L}-\boldsymbol{R}_B,\boldsymbol{s})$$

For finite cluster expansion, this method is obviously only approximate but we shall see that in the case of LiH, MgO and Si, the 1-electron properties converge towards the crystal ones, calculated with the Crystal95 program package[10] at HF level.

The CPM should be efficient, though approximate, for systems where electron delocalization effects remain moderate. As a molecular-like approach, it bypasses the summation over the first Brillouin zone: no periodicity of the system is needed. As such, it also permits investigations of defects, provided that an appropriate partition scheme for the 1RDM is available. Moreover, using electronic structure calculation codes such as Gaussian94[11] further allows for calculations with explicitly correlated wave functions. Besides, CPM turned out to be particularly efficient for molecular solids such as Ice $I_h$[12].

## III. Convergence of the CPM

First, we examine the case of MgO ionic clusters. Calculations are performed on clusters of 27 ions (figure 1, including third shell of neighbors), surrounded by point charges that ensure electrical neutrality. Basis-sets used here are these developed by Harrison and co-workers[13,14,15,16] (*i.e.* 8411G* for O and 8511G* for Mg). Differences obtained for the isotropic Compton profile between cluster and Crystal95 approaches are not visible at the scale of the plot (figure 2). CPM anisotropies also compare favorably: they further are in close agreement with experimental ones measured by Fluteaux *et al*[17,18] (see an example on figure 3). The relative error on the total electronic population is about 0.015%. Tests have also been performed with CaO clusters, for which both anisotropies and isotropic profiles have been found to be very close to Crystal95 ones.

$Mg^{2+}$ and $O^{2-}$ form relatively well closed-shell systems, the long-range electrostatic influence of which can be fairly recovered from point charges. Conversely, the case of LiH gives rise to strong overlaps occurring between distant anions. We therefore need to extend the cluster size up to the ninth shell of neighbors to achieve convergence. The small number of electrons involved enables such cluster calculations at HF level and with the basis-set recommended by Dovesi[19]. Clusters are again neutralized using point charges. We can check the convergence of the value $J(0)$, usually very sensitive to the level of theory chosen, for the [100] and [110] directions on table 1. The anisotropy $J_{[100]} - J_{[110]}$ still better reveals the slow convergence of LiH profiles (figure 4) with the cluster size. Mulliken populations obtained are compared in table 2, as well as kinetic energies (referring to one LiH cell) which differ from less than 0.05%.

The usual approach for cluster-based calculations on covalent crystals is to saturate periphery bonds with hydrogen atoms[20]. In the case of silicon, both Si-H distances and H orbital exponents can be optimized in order to provide quasi-exact density-matrix elements for central Si atoms (Si-H optimized distances vary around 1.6 Å, depending on the basis-set chosen). We have checked that an HF/8-41G calculation on $Si_5H_{12}$ partitioned cluster already result in static structure factors close to Crystal95 ones[21] (relative errors remaining less than 0.6%). The agreement becomes almost perfect when extending the cluster to $Si_{35}H_{36}$ (table 3).

## IV. Local correlation effects on ionic compounds

Comparing experimental and computed HF Compton profiles in MgO and LiH crystals, a significant isotropic discrepancy occurs. These effects remain with different experiments[22] and within different basis-set calculations[17,31] (see figures 5 and 6). Furthermore, the observed deviations to isotropic HF profiles show opposite trends for both compounds.

Assuming that correlation effects are mainly local[23] (*i.e.* intra-ionic), we investigate small clusters. Hence, the following correlation profiles $\Delta J_{Corr} = J_{Cluster/Corr} - J_{Cluster/HF}$ originate from clusters of 7 ions ($A_1^{\delta_A} B_6^{\delta_B}$-like clusters) embedded in a Madelung field and are compared to the observed deviation $\Delta J_{Exp} = J_{Exp} - J_{Crystal95/HF}$. Small clusters enable calculations at correlated level. Dovesis's basis-set has been extended with optimized *s* and *p* functions for Li and H, whereas we kept Harrisson's basis-set in the case of MgO. In each case, the contribution of periphery ions to the Compton profile is averaged, which will be hereafter reminded through the notation: $A_1^{\delta_A} \langle B_6^{\delta_B} \rangle$. The correlation profiles $\Delta J_{Corr}$ thus refer to either 4 or 20 electrons (respectively for LiH or MgO). We expect to catch part of anion-anion correlation effects through calculations on $A_1^{\delta+} B_6^{\delta-}$-like clusters.

For LiH, calculations are performed at QCISD[11] level on both $Li_1H_6$ and $H_1Li_6$ clusters, respectively centered on $Li^+$ and $H^-$. The two resulting correlation profiles appear on figure 7 and are compared to the experimental one[3]. The experimental deviation shown is multiplied by 0.525 in order to match the residual amplitude of deviations found by Bellaiche and Lévy[24], after correction for both impulse approximation and multiple scattering. Error bars refer to statistical errors.

Notice that the transfer of one electron from Li to H forms a paired couple, the correlation energy of which participates to the cohesion mechanism. The experimental cohesion energy was estimated[25] to 0.18 atomic units (a.u.) *vs*. 0.13 a.u. from HF calculation. The difference (0.05 u.a.) is of the order of the usually expected correlation energies for an isolated single pair of electrons[26] ($\approx 0.04$ a.u.). For a qualitative comparison, we estimated the correlation kinetic energy (QCISD − HF) of anionic electrons from a Mulliken partition: they were found to be 0.04 and 0.03 a.u. for $Li_1\langle H_6\rangle$ and $H_1\langle Li_6\rangle$ clusters, respectively. The difference might be attributed to electron correlation between anions. These values approach the expected one (which should not exceed 0.05 a.u., through the Virial theorem). In comparison, the experimental deviation to HF isotropic profile of LiH widely overestimates the correlation kinetic energy (this will be discussed in section V).

The theoretical correlation profiles show however the same trend as the experimental deviation, that is, $\Delta J_{Corr}(0) < 0$ and as such, are not characteristic of these of 2-electron free ions. Restoring the basis-set to the original Dovesi's one enables calculations on larger clusters, which confirmed this conclusion, although theoretical correlation profiles exhibit smaller magnitudes. In each case, correlation does not bring appreciable corrections to charge transfer. Rather, the correlation profiles found for LiH suggest a preferred angular correlation mechanism to be dealt with in a forthcoming paper[27]. Since angular correlation enables electrons to get closer to the nucleus, it reduces the size of the anion, which is consistent with the often-mentioned HF overestimation of anionic radii in LiH[5].

A similar approach has been used for MgO: we report on figure 8 the correlation profiles of $Mg_1\langle O_6\rangle$ and $O_1\langle Mg_6\rangle$ clusters calculated at MP2 level (using valence excitations only). The resulting correlation profiles bear again some resemblance with the experimental deviation, though we miss a large part of the correlation effects. Indeed, the limitation on the level of theory involved (MP2) underestimates the number of electrons at high values of momenta. Correlation kinetic energies are therefore widely underestimated: we found 0.48 and 0.32 a.u., respectively for $O_1\langle Mg_6\rangle$ and $Mg_1\langle O_6\rangle$. In comparison, the correlation energy for the embedded anion[4] *alone* was estimated to -0.36 a.u. (which should roughly correspond to an increase of 0.36 a.u. of the kinetic energy). The trend of both correlation profile and experimental deviation evoke this time the correlation profiles of free 2- or 10-electron systems[28]. Replacing Harrison's $O^{2-}$ basis-set with the cc-pV6Z one[29] (troncated after *p* orbitals for feasibility) further led to a similar correlation profiles, though larger in magnitude at low momenta. Note that experimental core profiles of MgO have been subtracted following Issolah's procedure[30]. Data have however not been corrected for multiple scattering, which could overestimate the experimental deviation amplitude (as suggested from experiments on LiH[24]).

A previous analysis of MgO Compton profiles, based on a repeated cluster model, has shown a weak electronic coupling between ions[31] together with a relatively high Slater exponent for anionic valence orbitals ($\zeta_{O,2sp}=1.8$). Such characteristics give the crystal the aspect of an assembly of *well* closed-shell systems, which is consistent with the correlation profiles we found. The fact that $\Delta J_{Corr}(q) > 0$ at small values of *q* is usually associated to a radial correlation mechanism (in the case of free 2- and 10-electron systems), which slightly widens electron distributions. Correlation between distant ions (Van der Waals effects) also participates in the cohesion. They were estimated by Doll and co-workers[4] to result in about 40% of the correlation energy participating to the cohesion of the crystal. These authors further estimated both contributions Mg-O and O-O as being about equal, which recommends including the second shell of neighbors for future estimations of correlated profiles.

Despite limitations on both cluster sizes and levels of theory, correlation profiles obtained confirm the observed experimental trends, that is, $\Delta J_{Corr}(0) < 0$ for LiH and $\Delta J_{Corr}(0) > 0$ for MgO. Notice that $\Delta J_{Corr}(q)$ is in each case positive at high momenta, which is consistent with the Virial theorem. An analysis of Mulliken contributions to correlation profiles has further shown that correlation effects of larger magnitude are mainly associated to the anions.

Taking advantage of the validity of the CPM in both **R** and **P**-space, we now analyze the influence of correlation on MgO structure factors. A comparison with experimental results requires taking into account thermal effects. To achieve this, the simplest method still relies on weighting each pseudo ion contribution $f_n$ by a Debye-Waller factor ($W_n$):

$$\langle F(\mathbf{Q}) \rangle = \sum_n f_n(\mathbf{Q}) W_n(\mathbf{Q}) e^{i\mathbf{Q}.\mathbf{R}_n} \quad \text{and} \quad W_n(\mathbf{Q}) = e^{-B_n \sin\theta^2 / \lambda^2} \tag{8}$$

$B_n$ denotes the standard thermal parameter (assumed isotropic) for ion $n$. Pseudo ion contribution $f_n$ are extracted from CPM. Experimental results chosen for comparison are the same as in references 32 and 33: the 6 smaller angle reflections issued from an electron diffraction experiment[34] whereas remaining reflections are obtained from X-ray diffraction[35] (the total number of reflections being 45). Given the small statistical errors (in all cases less than 1%), the weight attributed to each reflection is chosen constant. The minimized quantity is $\chi_n = \sum_{i=1,n} |F_{Obs}(\mathbf{Q}_i) - F_{Calc}(\mathbf{Q}_i)|^2$. Refining Debye-Waller factors yields similar results for both ions: we obtained $B_{Mg} = 0.336$ and $B_O = 0.303$, respectively from HF calculations on both $Mg_{13}O_{14}$ and $O_{13}Mg_{14}$ clusters (afterwards partitioned). These values remain within about 1% after correlation corrections and are further in very close agreement with those found in references 32 and 33. Results obtained for structure factors are shown on figure 9, before and after correlation corrections (corresponding to that applied for Compton profiles). They are further compared to both experimental and DFT ones. The DFT results issued from an original method developed by Cortona, which provides optimal physical parameters for ionic compounds[36]. The quality of the fit is further characterized by the factor $R_n = \sum_{i=1,n} |F_{Obs}(\mathbf{Q}_i) - F_{Calc}(\mathbf{Q}_i)| / \sum_{i=1,n} |F_{Obs}(\mathbf{Q}_i)|$. Both $O_1\langle Mg_6 \rangle$ and $Mg_1 \langle O_6 \rangle$ correlation corrections are consistent and significant: the $R_9$ factor ($\times 10^3$) changes from 8.91 (CPM/HF calculation) to 5.85 and 6.80 after correlation corrections issued from $Mg_1 \langle O_6 \rangle$ and $O_1 \langle Mg_6 \rangle$ calculations, respectively. These corrected values are closer to that calculated from Cortona's results: 6.63.

## V. Conclusions and perspectives

A cluster-based calculation, followed by an appropriate partition of the 1RDM, allows for an accurate evaluation of 1-electron properties: Compton profiles, structure factors and kinetic energies. The local character of pseudo ion contributions simplifies structure factor refinements, while keeping the same 1RDMs used for Compton profile calculations. The CPM was tested for generic materials like ionic and covalent crystals and can obviously be applied to molecular crystals like Ice $I_h$, for which it seems even more appropriate[12]. Since periodicity of the system is not required, the CPM, together with usual embedding techniques, should help in understanding the role of disorder effects. However, the main difficulty when dealing with defects remains to define an appropriate partition of the 1RDM.

Restriction to small clusters further allows for a fair description of local correlation effects on both Compton profiles (LiH and MgO) and structure factors (MgO). In each case, correlation effects are mainly associated to the anions. Subsequent correlation profiles show a qualitative agreement with the observed experimental trends, that is, $\Delta J_{Corr}(0) < 0$ for LiH and $\Delta J_{Corr}(0) > 0$ for MgO. Such features might invoke different correlation mechanisms[27].

Calculations must however be improved in order to achieve better comparisons with experimental deviations, especially at high momenta. Local correlation effects on Si crystals are currently investigated.

# Acknowledgments

We are pleased to express our gratitude to Pietro Cortona for providing us with theoretical data on MgO structure factors and for stimulating discussions.

CAPTIONS TO FIGURES

Figure 1: NaCl-like Clusters of 27 ions

Figure 2: Isotropic profiles of MgO. *Full line*: CPM result. *Large dashed*: Crystal95. *Small dashed*: relative differences magnified by 100. Relative errors remain less than 0.15% for any momentum value

Figure 3: Compton profile anisotropy of MgO ([100] − [110]). *Full line*: CPM. *Dotted*: Crystal95. Both theoretical anisotropies were convoluted with the experimental resolution. *Dots with statistical error bars*: experimental (from reference 17)

Figure 4: CPM convergence of profile anisotropy ([100] - [110]) of LiH.

Figure 5: Isotropic Compton profiles of LiH. *Dotted line*: experimental profile corrected for impulse approximation (from 3). *Full*: HF profile (calculated within Dovési's basis-set[19]) and convoluted for experimental resolution. *Dashed*: experimental deviation to convoluted HF profile

Figure 6: Isotropic valence Compton profiles of MgO. *Dotted line*: experimental profile (corrected for impulse approximation, from 17). *Full*: HF profile (calculated within basis-set of Harrison *et al.*), convoluted with the experimental resolution. *Dashed*: experimental deviation to convoluted HF profile

Figure 7: Isotropic correlation profiles derived from $Li_1H_6$ et $H_1Li_6$ clusters, after averaging the periphery ion contributions. *Full*: $Li_1\langle H_6\rangle$ contribution. *Dotted*: $H_1\langle Li_6\rangle$ contribution. *Dots* (with error bars): $\Delta J_{Exp} = J_{Exp} - J_{Crystal95/HF}$, from [3] and scaled by a 0.525 factor (see text)

Figure 8: Isotropic correlation profiles derived from $Mg_1O_6$ et $O_1Mg_6$ clusters, after averaging the periphery ion contributions. *Full*: $Mg_1\langle O_6\rangle$ contribution. *Dotted*: $O_1\langle Mg_6\rangle$ contribution. *Dots* (with error bars): $\Delta J_{Exp} = J_{Exp} - J_{Crystal95/HF}$, from (from 17)

Figure 9: Comparison of residual relative errors on MgO refined structure factors. *White squares*: CPM/HF + $O_1\langle Mg_6\rangle$ correlation corrections. *Black squares*: CPM/HF + $Mg_1\langle O_6\rangle$ correlation corrections. *Circles*: Crystal95/HF. *Grey triangles*: DFT (from Gillet and Cortona[32])

# FIGURES

Figure 1

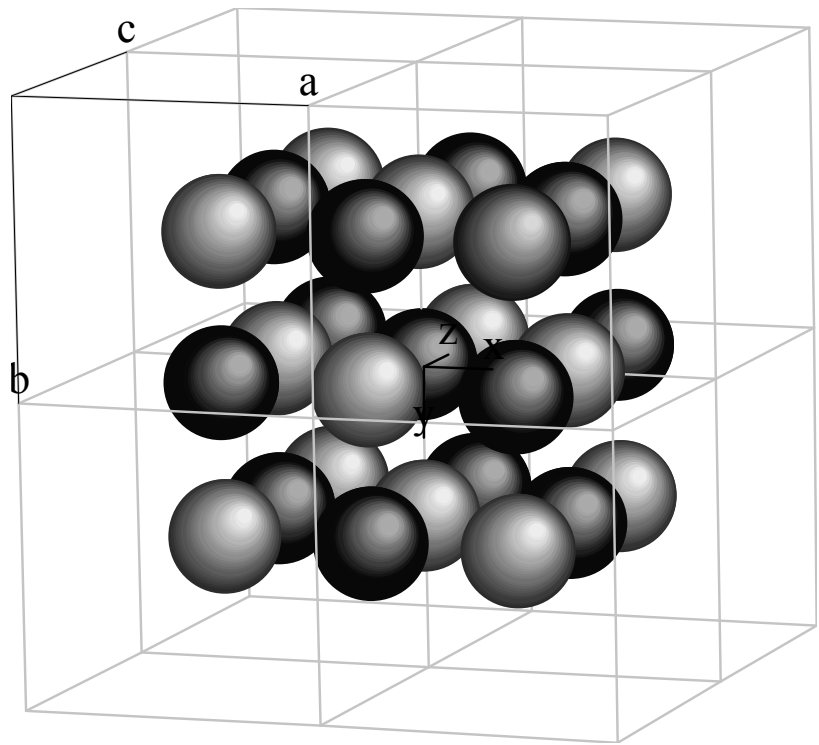

Figure 2

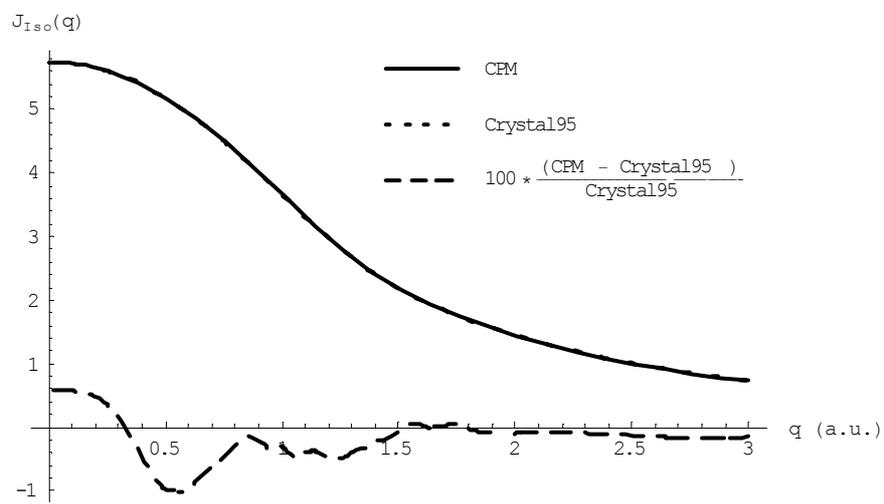

Figure 3

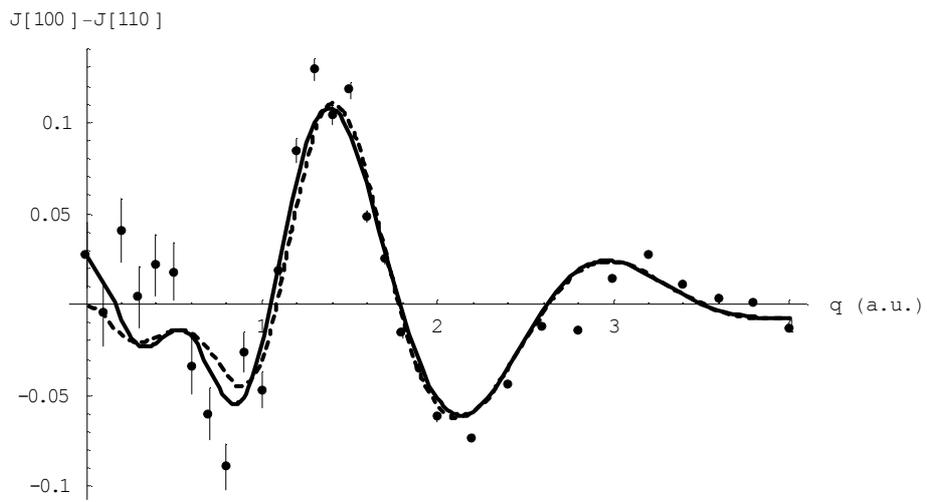

Figure 4

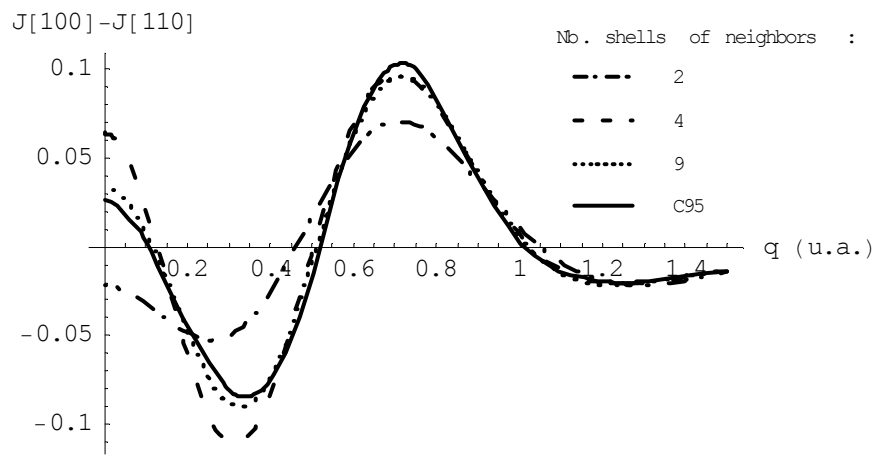

Figure 5

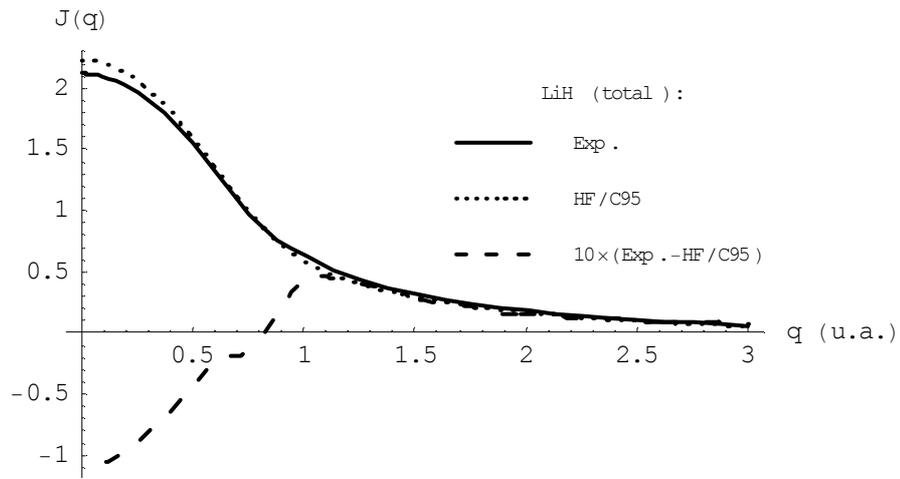

Figure 6

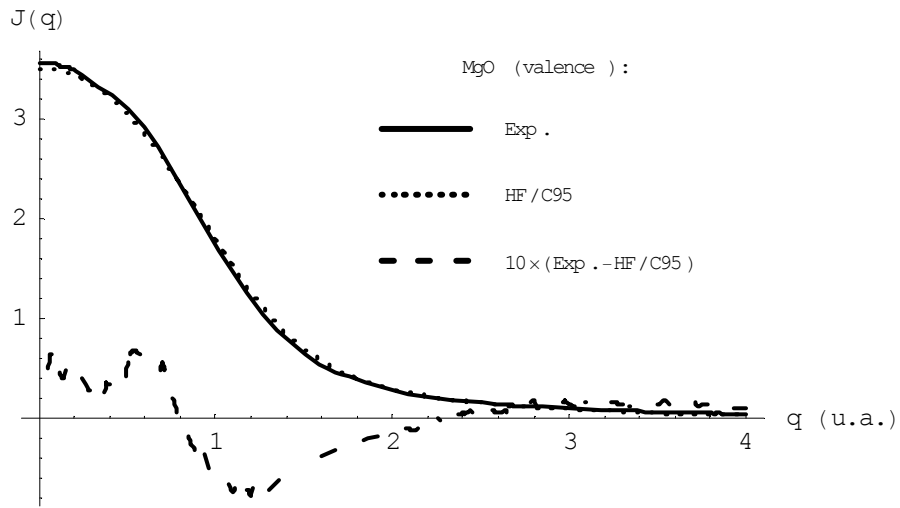

Figure 7

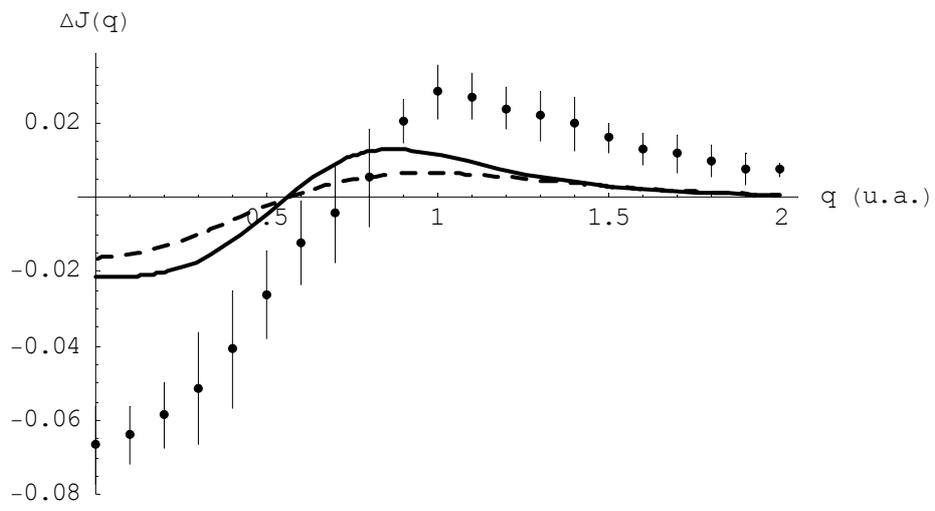

Figure 8

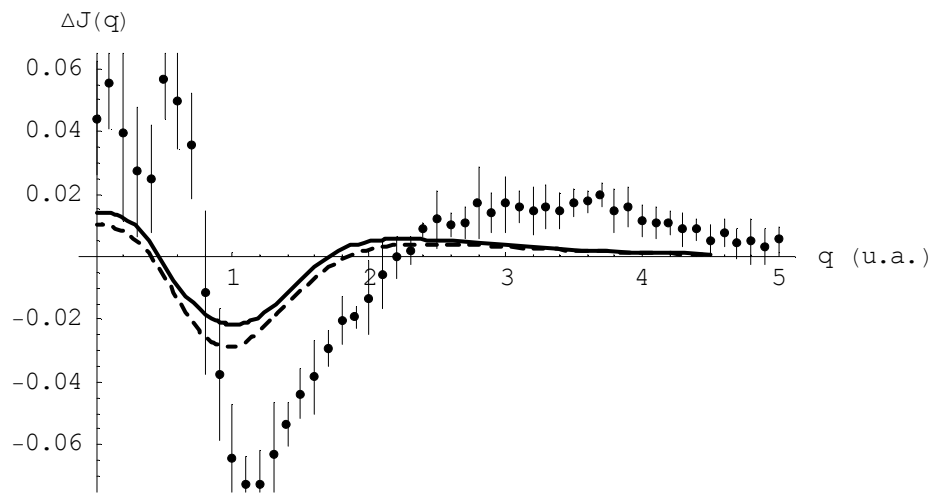

Figure 9

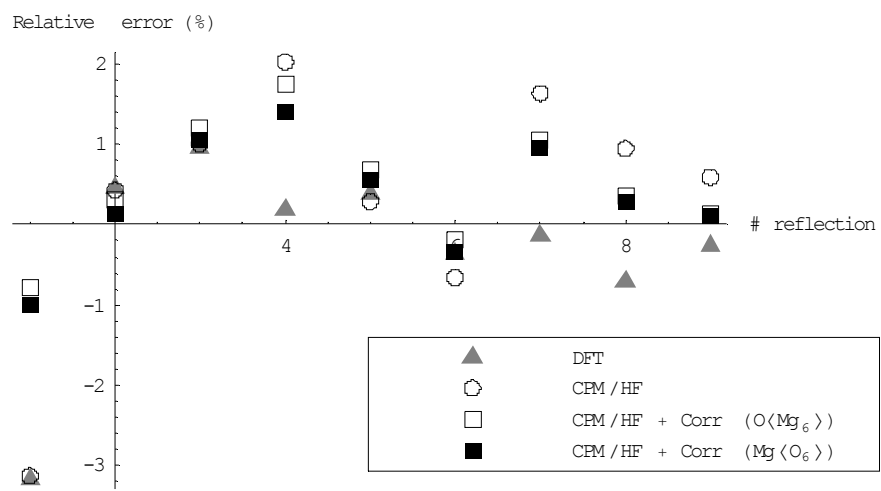

CAPTIONS TO TABLES

Table 1: Convergence of $J_{[100]}(0)$ and $J_{[110]}(0)$ for LiH CPM calculations
Table 2: Comparison of Mulliken populations and kinetic energies (a.u.) for LiH
Table 3: Static structure factors of Si estimated from Crystal95 and from a $Si_{35}H_{36}$ partitioned cluster (at HF/8-41G level). The accuracy is that delivered in output by Crystal95

TABLES

Table 1

| *Number of shells/Direction* | [100] (1$^{rst}$ neighbor direction) | [110] (2$^{nd}$ neighbor direction) |
|---|---|---|
| 0 | 2.60245 | 2.60245 |
| 1 | 2.31732 | 2.43376 |
| 2 | 2.13206 | 2.15404 |
| 3 | 2.15571 | 2.18046 |
| 4 | 2.24207 | 2.17809 |
| 9 | 2.24888 | 2.21568 |
| Crystal95 | **2.24310** | **2.21649** |

Table 2

|  | *H Population* | *Li$^+$ Population* | *Total Population* | *Total Kinetic energy* |
|---|---|---|---|---|
| *Crystal95* | 1.984 | 2.016 | 4.000 | 8.073 |
| CPM 9$^{th}$ shell | 1.979 | 2.020 | 3.999 | 8.069 |

Table 3

| h. | k. | l. | *|F(Q)|* Crystal95 | *|F(Q)|* $Si_{35}H_{36}$ *CPM* |
|---|---|---|---|---|
| 1. | 1. | 1. | 15.1 | 15.1 |
| 2. | 2. | 0. | 17.29 | 17.28 |
| 3. | 1. | 1. | 11.41 | 11.41 |
| 2. | 2. | 2. | 0.1777 | 0.1807 |
| 4. | 0. | 0. | 14.96 | 14.96 |
| 3. | 3. | 1. | 10.21 | 10.21 |
| 4. | 2. | 2. | 13.45 | 13.45 |
| 3. | 3. | 3. | 9.131 | 9.132 |
| 5. | 1. | 1. | 9.132 | 9.133 |

# References

[1] See for example Fulde P.: *"Electron Correlations in Molecules and Solids"*. Springer-Verlag, 2$^{nd}$ edn.(1993)

[2] See Parr R. G. and Yang W.: *"Density-Functional Theory of Atoms and Molecules"*. Oxford University Press (1989)

[3] Gillet J.M., Becker P., Loupias G. (1995), *Acta Cryst.* **A**, **51** 405

[4] Doll K., Dolg M., Stoll H., *Phys. Rev. B* **54** (1996) 13529

[5] Shukla A., Dolg M., Fulde P. *Phys.Rev.*B **60** (1999) 5211

[6] See for example the possibilities offered by Dovesi R. , Saunders V.R. , Roetti C. , Causà M. , Harrison N.M. , Orlando R. and C.M. Zicovich-Wilson, *CRYSTAL98 User's Manual*, Università di Torino (Torino,1998) and a sample of papers like these of Azavant P., Lichanot A., *Acta Cryst.* **A** 49(1993) 91 and Civalleri B., Casassa S., Garrone E., Pisani C. and Ugliengo P., *J. Phys. Chem.* **103**, 2165 (1999)

[7] See the road map proposed by Thakkar A. J. *et al.*: «On one-Matrices and related densities». *Density Matrices and Density Functionals*, eds., Erdahl R. M. and Smith V., Jr., 5-20, (1987) Riedel

[8] For a concrete analysis of the auto correlation function in insulators, see Patisson P., Weyrich W., *J. Phys. Chem. Sol.* **40** (1979) 40

[9] See a discussion for impulse approximation in Williams B: *"Compton Scattering"* (1977) Mac-Graw Hill Inc.

[10] Dovesi R. , Saunders V.R. , Roetti C. , Causà M. , Harrison N.M. , Orlando R. , Aprà E., Crystal95 User's Manual, University of Torino, Torino, 1996. See the documentation on line: http://www.ch.unito.it/ifm/teorica/crystal.html

[11] Gaussian 94, Revision D.4, Frisch M. J., Trucks G. W., Schlegel H. B., Gill P. M. W., Johnson B. G., Robb M. A., Cheeseman J. R., Keith T., Petersson G. A., Montgomery J. A., Raghavachari K., Al-Laham M. A., Zakrzewski V. G., Ortiz J. V., Foresman J. B.,  Cioslowski J., Stefanov B. B., Nanayakkara A., Challacombe M., Peng C. Y., Ayala P. Y., Chen W., Wong M. W., Andres J. L., Replogle E. S., Gomperts R., Martin R. L., Fox D. J.,  Binkley J. S., Defrees D. J., Baker J., Stewart J. P., Head-Gordon M., Gonzalez C., and Pople J. A., Gaussian, Inc., Pittsburgh PA, 1995

[12] Ragot S., Gillet J.M., Becker P.J. Submitted to Phys. Rev. B (11/2001)

[13] McCarthy M.I. and Harrison N.M., *Phys. Rev.* **B 49**, 8574 (1994)

[14] Towler M.D. et al. *Phys. Rev. B*. 52, 5375 (1995)

[15] Harrison M.and Saunders V.R., *J. Phys.:Cond.Mat.* **4**, 3873 (1992)

[16] Harrison N.M.et al. *J. Phys.: Cond. Mat.* **4**, L261 (1992)

[17] Fluteaux C., Gillet J.M., Becker P. (2000) *J Phys Chem Solids* **61** 369

[18] Notice that momentum density can be accessed and analyzed through Compton scattering (see Gillet J.M., Fluteaux C, Becker P., *Phys. Rev. B* **60** (1999) 2345) and that pair density function can in principle be observed through total X-Ray scattering (Meyer H., Müller T., Schweig A.. *Chem. Phys* **191** (1995) 213-222)

[19] H : 511G*, Li : 611G, R. Dovesi et al., *Phys. Rev.* **B 29**, 3591 (1984)

[20] See the "cluster" procedure included in Crystal95 programs

[21] The matrix element of the central Si atom have been afterwards duplicated and symmetrized in order to respect the centro-symmetry of the crystal

[22] The same trend is to be seen when using Aikala's measurements: Aikala O., Paakari T., Manninen S., (1982), *Acta Cryst. A*, **38** 155

[23] See ref. 4 for a comparison of intra and inter ionic estimated correlation energies in MgO and CaO

[24] See Bellaïche L. and Lévy B. Phys.Rev. B **54** (1996) 1575 and Bellaïche L. *Thesis*. Université Paris VI

[25] See O.L. Anderson, J. Phys. Chem. Solids. **27** (1966) 547 and reference 5

[26] Davidson E.R.: *"Reduced Density Matrices in Quantum Chemistry"*. Academic Press (1976)

[27] Ragot S., Gillet J.M., Becker P.J. In preparation

[28] Meyer H. Müller T., Schweig A. *J. Mol. Struct. (Theochem)* **360** (1996) 55

[29] The cc-pV6Z basis sets was obtained from the Extensible Computational Chemistry Environment Basis Set Database, Version 9/12/01, as developed and distributed by the Molecular Science Computing Facility, Environmental and Molecular Sciences Laboratory which is part of the Pacific Northwest Laboratory, P.O. Box 999, Richland, Washington 99352, USA, and funded by the U.S. Department of Energy. The Pacific Northwest Laboratory is a multi-program laboratory operated by Battelle Memorial Institue for the U.S. Department of Energy under contract DE-AC06-76RLO 1830. Contact David Feller or Karen Schuchardt for further information. E-mail: df_feller@pnl.gov. Web site: http://www.emsl.pnl.gov:2080/forms/basisform.html

[30] Issolah A., Levy B., Beswick A. and Loupias G.. *Phys. Rev. A* **38** (1988), pp. 4509-4512

[31] Fluteaux C. *Thesis*. Ecole Centrale Paris. (1999)

[32] Gillet J.M., Cortona P., *Phys. Rev. B* **60** (1999) 8569

[33] Gillet J.M., Becker P.J., Cortona P., *Phys. Rev. B* **63** (2001) 235115

[34] Zuo J.M., O'Keefe M., Rez P., Spence J.C.H., *Phys. Rev. Lett.* **78** (1997) 4777

[35] Destro R., Bianchi R, Gatti C, Merati F., *Chem. Phys. Lett* **186** (1991) 7

[36] Cortona P., *Phys. Rev. B* **46** (1992) 2008